\documentclass[aas_macros]{spie}  

\usepackage{amsmath,amsfonts,amssymb}
\usepackage{graphicx}
\usepackage{siunitx}
\usepackage{subcaption}
\usepackage[colorlinks=true, allcolors=blue]{hyperref}

\title{Gemini Planet Imager Observational Calibrations XIV: Polarimetric Contrasts and New Data Reduction Techniques}

\author[a,b]{Maxwell A. Millar-Blanchaer}
\author[c]{Marshall D. Perrin}
\author[d]{Li-Wei Hung}
\author[d]{Michael P. Fitzgerald}
\author[e]{Jason J. Wang}
\author[b]{Jeffrey Chilcote}
\author[e]{James R. Graham}
\author[f]{Sebastian Bruzzone}
\author[e]{Paul G. Kalas}
\author[]{and the GPI collaboration}

\affil[a]{Department of Astronomy \& Astrophysics, University of Toronto, 50 St. George St., Toronto, Ontario, Canada}
\affil[b]{Dunlap Institute for Astronomy \& Astrophysics, University of Toronto, 50 St. George St., Toronto, Ontario, Canada}
\affil[c]{Space Telescope Science Institute, Baltimore, MD 21218, USA}
\affil[d]{Department of Physics \& Astronomy, University of California, Los Angeles, CA 90095, USA}
\affil[e]{Astronomy Department, University of California, Berkeley; Berkeley CA 94720, USA}
\affil[f]{Department of Physics and Astronomy, Centre for Planetary Science and Exploration, the University of Western Ontario, London, ON N6A 3K7, Canada}

\authorinfo{Send correspondence to Maxwell A. Millar-Blanchaer\\
E-mail: maxmb@astro.utoronto.ca\\
}

\begin{document} 
\maketitle


\begin{abstract}
The Gemini Planet Imager (GPI) has been designed for the direct detection and characterization of exoplanets and circumstellar disks. GPI is equipped with a dual channel polarimetry mode designed to take advantage of the inherently polarized light scattered off circumstellar material to further suppress the residual seeing halo left uncorrected by the adaptive optics. We explore how recent advances in data reduction techniques reduce systematics and improve the achievable contrast in polarimetry mode. In particular, we consider different flux extraction techniques when constructing datacubes from raw data, division by a polarized flat-field and a method for subtracting instrumental polarization. Using observations of unpolarized standard stars we find that GPI's instrumental polarization is consistent with being wavelength independent within our errors. In addition, we provide polarimetry contrast curves that demonstrate typical performance throughout the GPIES campaign.
\end{abstract}



\section{INTRODUCTION}
\label{sec:intro}

The Gemini Planet Imager (GPI) is a high-contrast instrument on the Gemini South 8-m telescope designed for the direct detection and characterization of Jupiter-like planets and dusty debris disks around young nearby stars\cite{Macintosh2014}. Its optical design combines an extreme adaptive optics system\cite{Poyneer2014} and an apodized-pupil Lyot coronagraph\cite{Soummer2011} with a lenslet-based integral field spectrograph (IFS)\cite{Larkin2014}. GPI also includes a polarimetry sub-system composed of a rotatable half-wave plate (HWP) modulator and a Wollaston prism analyzer that when deployed replaces the spectrograph's prism \cite{Perrin2015}. The Wollaston prism disperses incident light such that each lenslet produces two spots of orthogonal linear polarization on the detector. For a detailed description of GPI's polarimetry mode we direct readers to Perrin et al. (2015)\cite{Perrin2015}. In addition to being available to the Gemini community as a standard observing mode, GPI's polarimetry mode is currently being used to carry out scattered-light ($H$-band) debris disk observations as part of the GPI Exoplanet Survey (GPIES). Target stars with known infrared excess are observed with a short `snapshot' observing sequence and stars with previously resolved disks, or disks discovered in a snapshot are observed for an hour long sequence. GPI is also being used to carry out a detailed study of debris disk composition as part of a Gemini Large and Long Program (PI: Christine Chen). However, the results of the large program study are not discussed in this work. 

Here we report on recent updates to the data reduction process and summarize typical performance of the polarimetry mode during the GPIES survey, building upon the early characterization work presented by Wiktorowicz et al. (2014)\cite{Wiktorowicz2014SPIE} and Perrin et al (2015)\cite{Perrin2015}. Throughout this work we will attempt to quantify performance improvements based on the 5-sigma contrast measured in the total linear polarized intensity, $P=\sqrt{Q^2+U^2}$.

In Section~\ref{sec:contrast} we summarize the reduction of polarimetry data and describe the method used to calculate contrast. In Sections~\ref{sec:extraction} and~\ref{sec:flatfielding} we discuss the methods used to convert the raw data to a polarization datacube and the use of polarized flat fields, respectively. We present observations of unpolarized standard stars that we use to assess GPI's instrumental polarization in Section~\ref{sec:instpol}, where we also present the current method used to subtract the instrumental polarization. In Section~\ref{sec:GPIES} we present the typical polarized intensity contrasts achieved in the GPIES survey, followed by some concluding remarks in Section~\ref{sec:conclusions}. 


\section{Data Reduction and Contrast Measurement in Polarimetry Mode}
\label{sec:contrast}

A standard polarimetry mode observation sequence involves taking images with the HWP at positions of 0, 22.5, 45 and 67.5 degrees, making GPI sensitive to linear polarization (Stokes Q and Stokes U). Data are reduced using the GPI Data Reduction Pipeline (DRP) which converts raw data into analysis-ready datacubes, where each step in the reduction is known as a `primitive'\cite{Maire2012, Perrin2014SPIE}. We briefly summarize the relevant data reduction steps here. 

In a typical reduction a raw data frame is first dark subtracted, corrected for bad pixels and cleaned for correlated detector noise. The raw data are then converted into a three-dimensional ``polarization datacube" where the first two dimensions are spatial dimensions and the third dimension holds two orthogonal polarization states. Therefore each spatial pixel (``spaxel") has two flux values. The polarization datacubes are then cleaned for bad pixels. At this point the position of the occulted star (behind the focal plane mask) is estimated using a radon-transform based method\cite{Wang2014} that relies on GPI's fiducial 'satellite spots', four replicas of the stellar PSF that are imprinted on the focal plane by a grid located in the apodizer. Each polarization cube can then be summed to obtain a total intensity image or differenced to obtain the linear polarized intensity with an orientation defined by the position of the HWP during the observation (stored in the FITS header of each file). A double differencing routine is then applied to all the polarization datacubes that serves to remove any bias introduced by non-common path errors for each lenslet. A sequence of polarization datacubes can be combined into a Stokes datacube, where the third dimension holds a Stokes vector ([I,Q,U,V]) at each spatial location, by solving a set of equations that describe the expected response of the instrument to incident polarized light given the known HWP and parallactic angles for each frame. Finally, the Stokes cube can be transformed to the radial Stokes convention ($[I,Q,U,V]\Rightarrow[I,Q_r,U_r,V]$) where a pixel in the $Q_r$ frame holds linear polarized intensity that is oriented either parallel (negative values) or perpedicular (positive values) to a line connecting the pixel to the central star's location. The $U_r$ frame holds polarization oriented $\pm45^\circ$ from that line. Under the assumption that the measured polarization is due to single scattering off relatively small dust grains, then all the flux should be in the $Q_r$ frame and the $U_r$ frame should just contain noise. 

The metric we use to quantify performance in polarimetry mode is the 5-sigma polarized intensity contrast, which we have developed  to be as similar as possible to the contrast measured in spectroscopy mode. In GPI's spectroscopy mode the final data product is a spectral datacube, where the third dimension holds spectral information. For a single wavelength slice, the 5-sigma point source contrast is measured at each angular separation as five times the standard deviation of the pixel values in an annulus around the star with that same angular separation. This value is then divided by the average peak satellite spot brightness of the four satellite spots in that wavelength slice and then multiplied by the known peak-satellite-spot-brightness-to-stellar-flux ratio to obtain the contrast relative to the stellar flux. 

In polarimetry mode the entire bandpass is seen in each polarization datacube and the satellite spots appear as elongated smears rather than as replica PSFs. Instead of measuring the peak satellite spot brightness, as is done in spectroscopy mode, we measure the total flux of the satellite spots using a DRP primitive called ``Measure Satellite Spot Flux in Polarimetry", that saves the flux information in the FITS header of the polarimetric datacube\cite{Hung2016SPIE}. The equivalent satellite spot peak brightness is estimated by multiplying the average of the total satellite spot fluxes of the four spots by a conversion factor. We assume a Gaussian PSF and the conversion factor is then the ratio between the the peak and total flux in a 2D Gaussian function. Thus, we are calculating the peak flux of a theoretical Gaussian PSF whose total flux is equal to the flux measured in the elongated satellite spots. The contrast is then calculated in the same manner as in spectroscopy mode, by dividing the standard deviation in concentric annuli by the equivalent satellite spot peak brightness and then multiplying by the peak-satellite-spot-brightness-to-stellar-flux ratio to obtain the contrast in units of stellar flux. The 5-sigma contrast is then calculated by multiplying by a factor of 5. 

To date no direct calibration of GPI's total-satellite-spot-flux-to-stellar flux ratio has been carried out in polarimetry mode. This is due in part to the difficulty of finding objects that are bright enough for the AO system to close control loops, but do not saturate the detector in non-coronagraphic observations when the Wollaston prism is in place (though a planned upgrade that includes installing an ND filter may solve this problem in the near future). While a direct measurement of the flux ratio is currently unavailable, laboratory and on-sky tests comparing the total satellite spot fluxes between spectroscopy mode and polarimetry modes for the same target indicate that the satellite spots in both modes receive the same amount of flux, implying that the flux ratio is the same\cite{Hung2016SPIE}.

\begin{figure}
\begin{center}
\includegraphics[width=0.7\columnwidth]{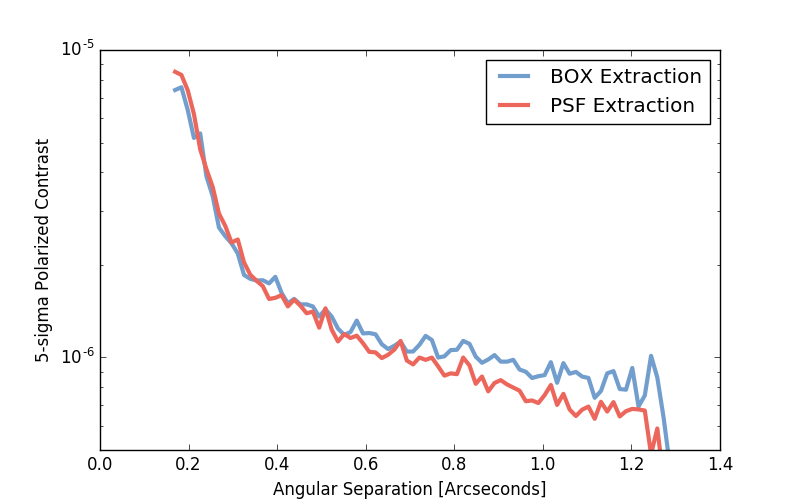}
\end{center}
\caption[Polarimetric Contrasts for Different Datacube Assembly Methods]{5-sigma polarized contrast for the observations of the unpolarized standard star HD 118666 using both a BOX and PSF extraction method. The PSF extraction method shows improvement over the BOX method outside of 0.3 arcseconds where the polarized contrast is dominated by photon and read noise. 
\label{fig:box_psf}}
\end{figure}

In polarization datacubes, contrast can be measured in total intensity as well as in polarized intensity (the difference of the two polarization states). In Stokes datacubes, the total satellite spot flux is estimated by taking the average value of the total satellite spot fluxes in all of the constituent polarization datacubes. Contrast can then be measured in any of the Stokes vector individual states (I, Q, U, or V), or in combinations of them, such as the polarized intensity ($P = \sqrt{Q^2+U^2}$). The DRP primitive ``Measure Contrast in Pol Mode" reports the total intensity and polarized intensity contrasts of both polarization datacubes and Stokes datacubes. The primitive gives the option to display the contrast immediately in a plot window and/or save the contrast information to a FITS file. Note that in spectroscopy mode, because we are typically searching for point sources, a high-pass filter can sometimes be applied to the data before measuring the contrast. Throughout this work we do not apply a high-pass filter to polarimetry mode data. 

Many of GPI's polarimetry mode targets are extended objects, such as protoplanetary or debris disks. The ability to detect extended objects above the noise floor is enhanced relative to point sources, by the fact that coherent structure can be extended across many pixels. The true detection threshold will then depend on the surface brightness and the angular extent on the sky of a given target. Thus, the point source contrast as defined in this section may somewhat underestimate the achievable detection limit for extended sources. However, as it is standard in the field, we opt to continue to use the 5-sigma point source contrast as the metric against which we test our new reduction techniques and report GPI's sensitivity.  


\section{Assembling Datacubes from Raw Data}
\label{sec:extraction}

A key step in the reduction of GPI data is the conversion from raw data to datacube. After the raw data has gone through dark subtraction, bad pixel correction and the subtraction of correlated detector noise, it is converted to a polarization datacube using a DRP primitive called ``Assemble Polarization cube". The flux in each lenslet spot is measured either by summing a square aperture centered on the lenslet, known as BOX extraction or summing via a weighted PSF. The location and morphology of each lenslets' two spots are predetermined using a polarization flat-field image obtained using Gemini's Facility Calibration Unit (GCAL). Each lenslet in the flat-field image is fit to a 2D Gaussian function and the best fit parameters, including x and y widths, tilt and center location, are saved in the polarization calibration file, known as a `polcal'. 

\begin{figure}
\begin{center}
\includegraphics[width=0.4\columnwidth]{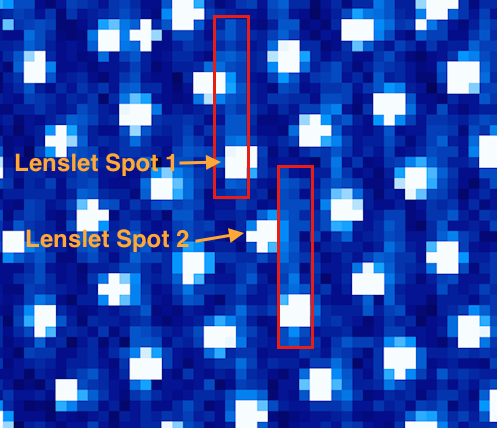}
\end{center}
\caption[Persistence in a Raw Detector Image]{A raw detector image from a polarization observation showing persistent spectra from a previous spectroscopic observation. Two spectra are highlighted in red (though more are apparent in the image) and the labels indicate the two orthogonally polarized spots produced by the Wollaston prism for one lenslet. The persistent spectra are aligned such that the flux measured from Lenslet Spot 1 receives a stronger bias than that measured for Lenslet Spot 2, resulting in an artificial polarization signal. 
\label{fig:spec_pol_alignment}}
\end{figure}

An analysis of observations of the unpolarized standard star HD 118666, presented in Perrin et al. (2015) \cite{Perrin2015}, demonstrated that for an 8 minute observation sequence, the polarized intensity contrast is limited by photon noise and read noise outside of $\sim$0.3 arcseconds when using the BOX extraction. The analysis was based on observations taken as part of GPI commissioning on March 24, 2014 and consisted of eight 60s integrations, with the HWP rotating by 22.5 degrees between each observation. Since that time we have included in the ``Assemble Polarization cube" primitive the option to use a weighted PSF extraction, where each pixel's weight is determined by its relative contribution to the lenslet PSF, which is assumed to be the best fit Gaussian for that lenslet as stored in the polcal calibration file. This has the effect of reducing photon and read noise of each polarization spot measurement in the extracted polarization datacube, because pixels where the relative noise is high get down-weighted in the sum. An added advantage of the weighted PSF technique is that bad pixels in the raw data can be masked out and the surrounding pixels can still be used to provide an estimate of the total flux in each lenslet spot. When using the BOX method bad pixels either have to be masked out or included in the sum, which can result in spurious values for spots that contain bad pixels.

In Figure~\ref{fig:box_psf} we present the polarized intensity contrast curve from the final Stokes cube of the HD 118666 dataset with both the BOX and PSF extraction methods. We find that the PSF extraction method improves the polarized intensity contrast starting at $\sim0.3''$, where the data are limited by photon and read noise in the BOX extraction. While the improvement can be as small as 5\% at 0.5 arcseconds, it reaches over 20\% at larger separations. As a result, weighted PSF extraction is now the default option in the current version of the DRP and is used exclusively in the remaining sections of this work. Establishing the relative contributions of different noise sources using the PSF extraction technique has been left for future work. 


\subsection{Persistence}
\label{sec:persistence}

Persistence, also known as latency, occurs in detectors when electrons previously freed by incident light, get trapped in the detector crystal lattice. When the detector is read out, the trapped electrons are indistinguishable from newly released electrons and cause a bias in the raw image. These electrons are not expelled when the detector is reset, but instead remain trapped and decay as a function of time. Persistence in the GPI H2RG can be as high as 21 e$^{\text{-}}$/s; calculated by taking a 60s exposure immediately after a saturating exposure.

\begin{figure}
\begin{flushleft}
\includegraphics[width=\columnwidth]{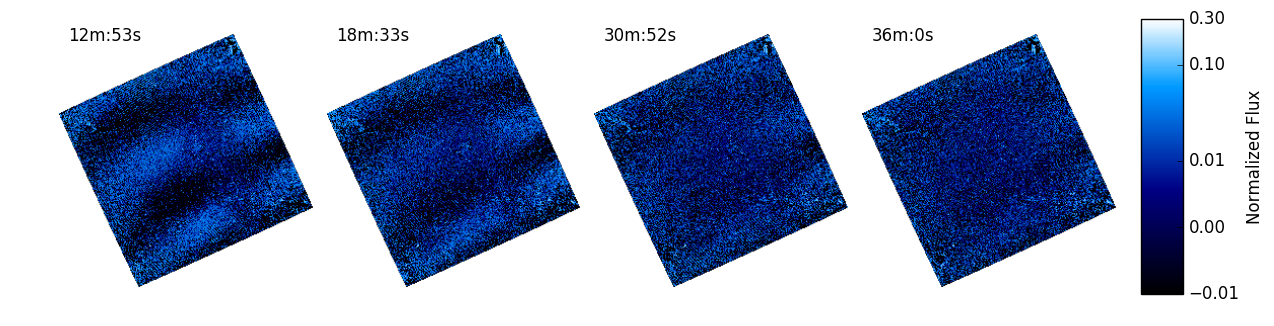}
\end{flushleft}
\caption[Persistence Over Time in a Polarization Datacube]{A time series of normalized difference images from four polarimetry datacubes taken from the same observation set (60s exposures), immediately following a spectroscopic observation of the same target. Persistence manifests as the thick vertical bar structure seen clearly in the first two images and diminishes to negligible levels by the fourth frame. The time labels indicate the time between the end of the last spectroscopic observation and the depicted polarimetry frame. 
\label{fig:persistence_imgs}}
\end{figure}

In GPI observations persistence is most apparent in polarimetry observations, due to the differential nature of the measurement. Though the fractional value of the persistent flux relative to a lenslet spot's flux may be small, because the polarization measurement relies on the difference in flux between the two spots, the bias can have a significant effect. This effect is strongest when polarimetric observations are taken immediately following a spectroscopic observation. Persistent spectra from the previous observation can coincide with the locations of one or both of the two polarization spots (e.g. Figure~\ref{fig:spec_pol_alignment}). The result is a polarization bias that depends on the strength of the persistence and the exact alignment of the spectrum on the two polarization spots. The relative alignment of the spectrum and the polarization spots changes across the detector resulting in a spatially dependent polarization bias (Figure~\ref{fig:persistence_imgs}). Because the strength of the polarization bias changes with time it can not be fully removed during the double differencing procedure and can masquerade as polarized emission.

Some detectors have shown a persistence which is a non-linear function of the total flux accumulated during an integration  incident flux, making the effects of persistence worse for brighter stars\cite{Long2012}. In our experience in the GPIES campaign we have found that back to back spectroscopic/polarimetric observations for the brightest stars ($I < 5$) can result in significant amounts of persistence lasting over 20 minutes after the polarimetric observations have begun (Figure~\ref{fig:persistence_imgs}). Detailed characterization of persistence is difficult because of the complicated physics involved in understanding detector crystalline structure. Our efforts in understanding the effects of persistence including detailed modeling and methods to mitigate its effects are ongoing. 


\section{Flat Fielding}
\label{sec:flatfielding}

\begin{figure}
\begin{center}
\includegraphics[width=0.9\columnwidth]{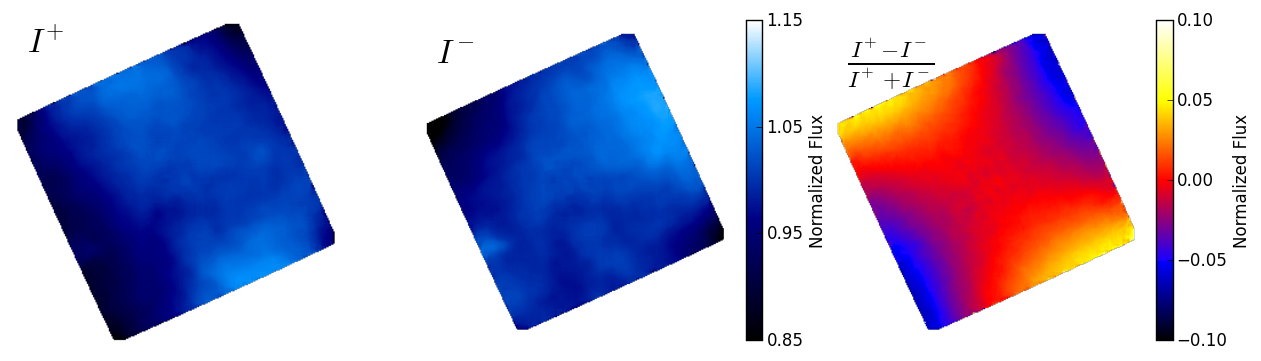}
\end{center}
\caption[Polarized Flat-Field Signal]{Images of the two orthogonal polarization slices (left, center) and their normalized difference (right) from a low-pass filtered GCAL flat field polarization datacube. The quadrupole pattern seen in the normalized difference is present in both GCAL flat field data, as well as on-sky observations. Each individual slices has been normalized by its median value. 
\label{fig:gcal_flatfield}}
\end{figure}

\begin{figure}
\begin{center}
\includegraphics[width=0.7\columnwidth]{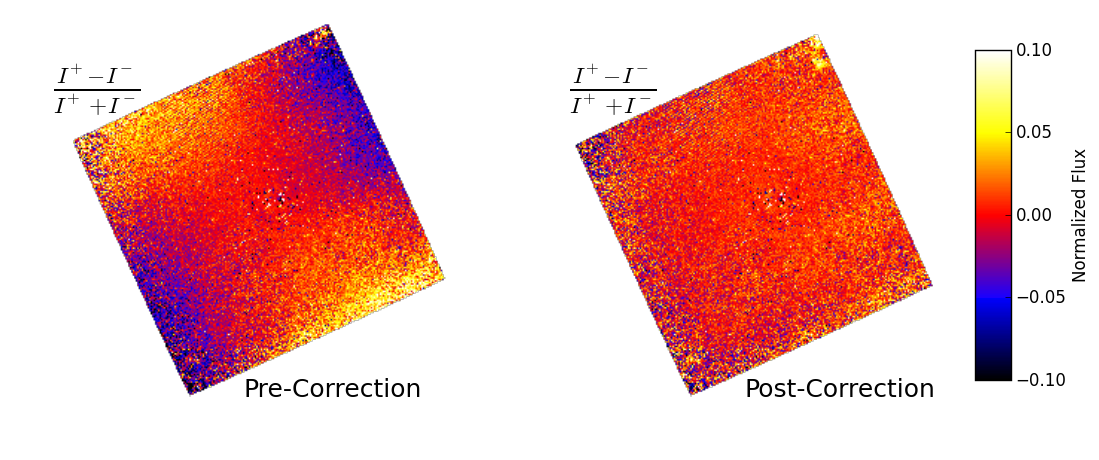}
\end{center}
\caption[Polarized Flat-Field Correction in a Polarization Datacube]{The normalized difference image of a polarization datacube from the HD 118666 dataset before (left) and after division by GCAL low spatial frequency flat field shown in Figure~\ref{fig:gcal_flatfield} (right). Dividing by the flat field successfully removes the polarization bias. Both images are shown with the same color stretch. 
\label{fig:podc_before_after}}
\end{figure}

GPI's polarization datacubes are subject to a polarization bias signal that is proportional to the incident flux and changes across the field of view (Figure~\ref{fig:gcal_flatfield} and Figure~\ref{fig:podc_before_after}). The flat field resembles a quadrupole pattern where the strength of the bias is strongest in the four corners, ranging from about $-5\%$ to $5\%$. This signal appears in polarization observation of both the Gemini GCAL calibration unit and on-sky targets. The bias does not change with HWP position, indicating that its origin is downstream of the HWP in the optical train. 

\begin{figure}
\begin{center}
\includegraphics[width=\columnwidth]{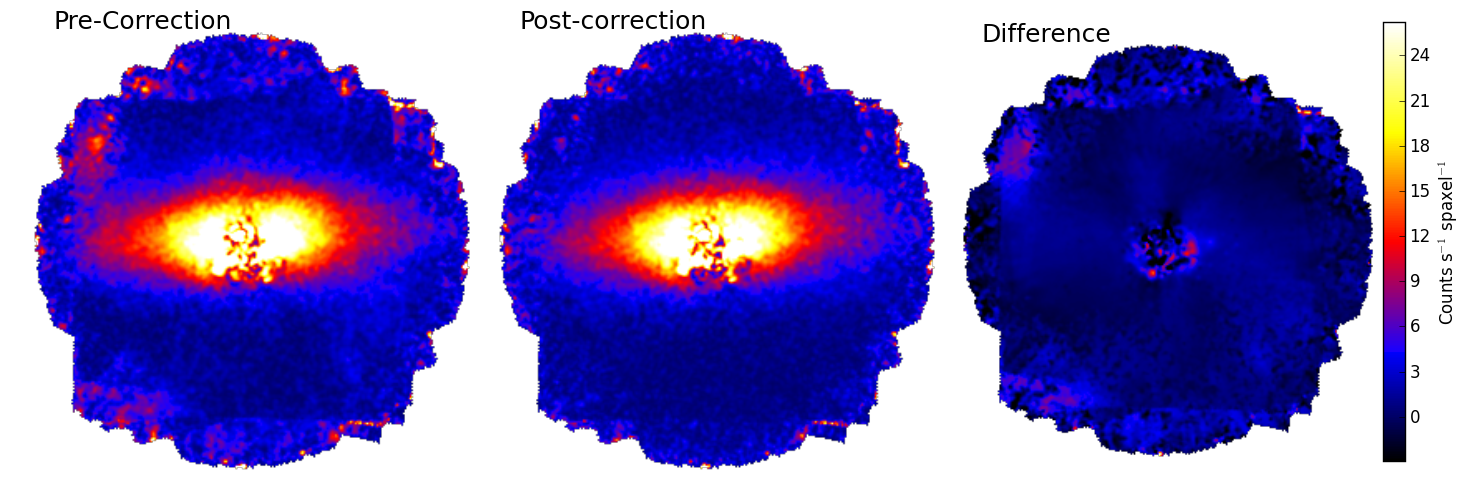}
\end{center}
 \caption[Polarized Flat-Field Correction in an Extended Observation Sequence]{Linear polarized intensity images of the $\beta$ Pictoris disk before (left) and after (center) flat field correction using a low spatial frequency polarized flat field. The disk images have been rotated so that the outer disk, with a position angle of $29^\circ$ is horizontal, as in Millar-Blanchaer et al. (2015)\cite{MillarBlanchaer2015}, where this data was first presented. The difference between the left and the center panel can be seen on the right. The correction affects mostly the outer regions, but in this case the inner regions have mild corrections (relative to the local flux levels) as well.
\label{fig:stokesdc_before_after}}
\end{figure}

This signal can be compensated for by dividing each polarization datacube by a ``low spatial frequency polarization flat field' datacube, a spatially filtered polarization flat field datacube (Figure~\ref{fig:gcal_flatfield}). Polarization flat field datacubes are created from observations of the GCAL Quartz-Halogen lamp, and are assembled into polarization datacubes in the same manner as standard polarization observations. Each 2D slice of the cube (i.e. one orthogonal polarization slice) is then normalized by dividing by its mean.  The flat field datacube is then filtered spatially using a Fourier transformed based-filter. This has been implemented in the DRP as a standard calibration recipe template called ``Create Low Spatial Frequency Polarized Flat-Field" and is available in DRP versions 1.4 and higher. This flat can be divided from a polarization datacube using a primitive called ``Divide by Low Spatial Freq. Polarized Flat Field". Figure~\ref{fig:podc_before_after} displays a polarization datacube before and after dividing by a low spatial frequency polarized flat-field. We have found that dividing by a polarization datacube that has not been smoothed results in a significant number of bad lenslets being artificially injected into the datacube.

In combined Stokes datacubes the flat-field bias can manifest as added noise at larger separations. For example, Figure~\ref{fig:stokesdc_before_after} displays a linear polarized intensity image of the $\beta$ Pic debris disk, originally presented by Ref.~\citenum{MillarBlanchaer2015}, before and after flat field correction. Significant noise can be seen at the edge of the field.  When each polarization datacube is divided by a low spatial frequency flat field before being combined into a Stokes cube the noise is largely eliminated. Because this signal is a multiplicative function of the input flux, it cannot be fully compensated for by the double differencing algorithm, which compensates for static bias offsets in each lenslet. However, for short sequences (e.g. 10-minute) where the PSF is relatively stable, the double differencing will compensate for most of the signal. For example, in the HD 118666 observations, the flat-field correction provides negligible improvement over the double differencing algorithm alone. Nonetheless, the flat field correction is now included as a standard primitive in the DRP recipes handling polarization datacubes. 


\section{Instrumental Polarization}
\label{sec:instpol}

\begin{figure}
\begin{center}
\includegraphics[width=0.9\columnwidth]{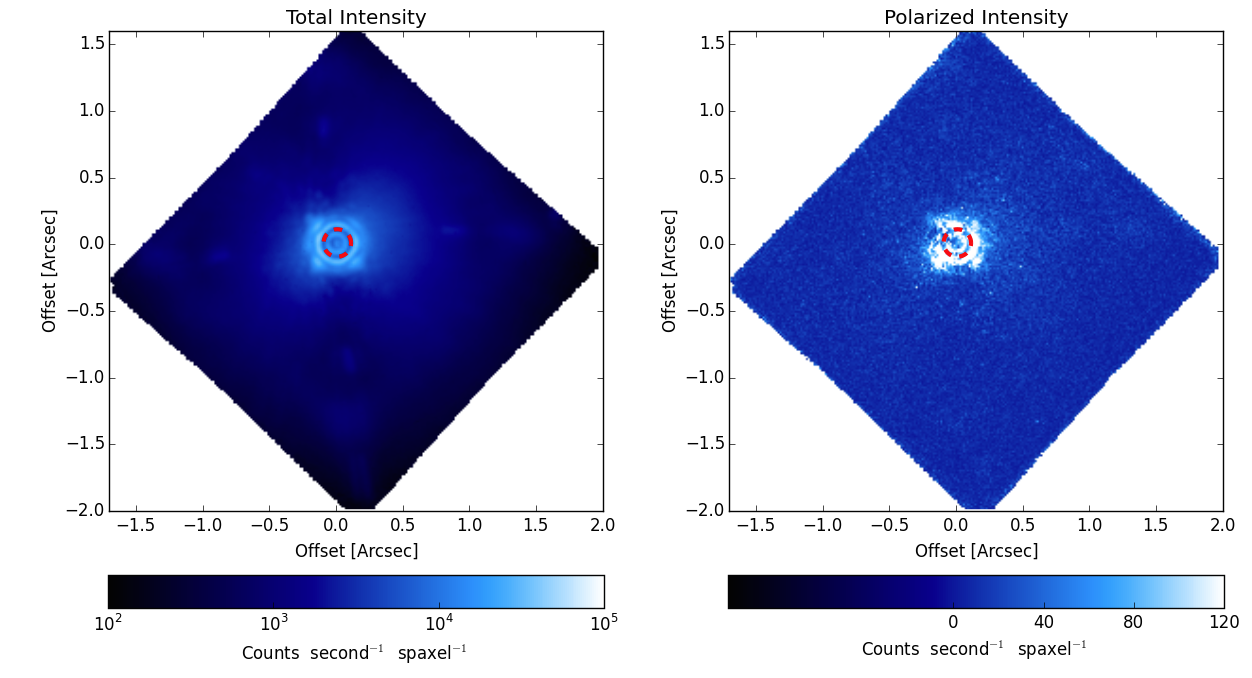}
\end{center}
\caption[GPI Total Intensity and Linear Polarized Intensity Images Demonstrating Instrumental Polarization.]{GPI $H$-band total intensity and linear polarized intensity images of HD 118666. The linear polarized intensity is predominantly due to instrumental polarization and its strength is proportional to the total intensity at a given location. The red dashed circle in each image denotes the angular extent of the $H$-band focal plane mask. 
\label{fig:ip_demo}}
\end{figure}

\begin{figure}
\begin{center}
\includegraphics[width=\columnwidth]{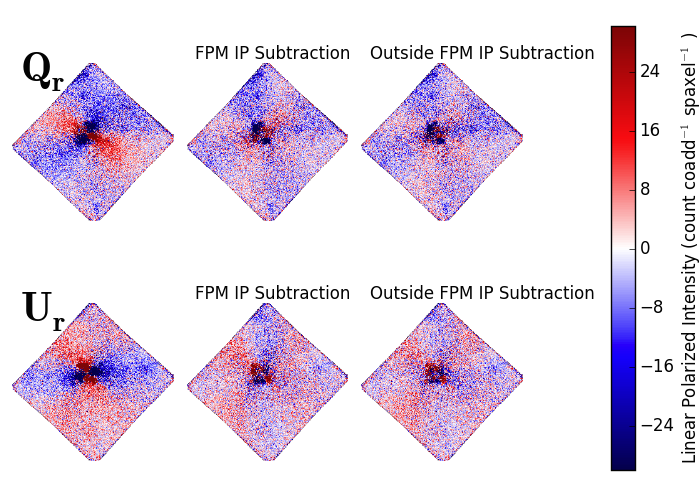}
\end{center}
\caption[Instrumental Polarization in the $Q_r$ and $U_r$ bases]{$Q_r$ and $U_r$ images of the GPI HD~118666 $H$-band observations. The first column of images shows the reduced cubes with no instrumental polarization (IP) subtraction. The instrumental polarization appears as a quadrupole pattern in both the $Q_r$ and $U_r$ images, offset $45^\circ$ from each other. The second and third column display the results of subtracting the instrumental polarization as measured from behind the coronagraph FPM and just outside the FPM, respectively. When subtracting instrumental polarization measured from outside the FPM, slight gains are made in the inner regions (most readily seen in the $Q_r$ image). A faint residual octopole can be seen in all of the IP-subtracted $Q_r$ and $U_r$ images. Understanding and compensating for this residual is a work in progress. 
\label{fig:qrur_ip_sub}}
\end{figure}

Polarization induced by optics upstream of GPI's HWP, or instrumental polarization, can masquerade as astrophysical signal since it also modulates with the rotation of the HWP. In reduced GPI polarization images instrumental polarization manifests as a polarization signal that is proportional to the residual total intensity of the stellar PSF at any given location (Figure~\ref{fig:ip_demo}) and whose position angle is constant across the frame. If a Stokes cube is converted to the radial convention, then the instrumental polarization manifests as a quadrupole pattern due to the constant position angle (e.g. Figure~\ref{fig:qrur_ip_sub}). The following two subsections describe our efforts to characterize GPI's instrumental polarization at multiple wavelengths and the current method used to subtract it from polarization datacubes. 

\subsection{Unpolarized Standard Stars}
In this section we present an analysis of GPI observations of several unpolarized standard stars, which we use to characterize GPI's instrumental polarization. This work builds upon the initial work carried out by Wiktorowicz et al. (2014)\cite{Wiktorowicz2014SPIE}, who leveraged a large amount of field rotation in coronagraphic observations of $\beta$ Pic to separate astrophysical signal from instrumental polarization (which keeps a constant position angle over time) to measure the $H$-band instrumental polarization to be $0.4354\pm0.0075\%$. Here we present measurements of the instrumental polarization in all 5 of GPI's broadband filters ($Y$, $J$, $H$, $K1$ and $K2$). 

We observed three unpolarized standard stars, HD 82386, HD 99171, and HD 210918, in GPI's direct mode (i.e. without a focal plane mask) as polarimetric calibrators during several of GPI's commissioning runs in 2014 (Table~\ref{tab:unpol}). To avoid saturating the detector within the minimum exposure time, the AO loop was set to open. In all exposures the tip/tilt loop was open, except for the last eight images of HD 210918. Each observation sequence consisted of 4 images where the HWP was rotated by $22.5^{\circ}$ in between each exposure. We reduced the raw data to polarization datacubes with the GPI DRP, using the methods described in Section~\ref{sec:intro}. Because we were observing in direct mode, the AO loop was open during these observations, and the star position and the PSF shape varied from exposure to exposure. We therefore opted to carry out our analysis on the total intensity in each individual polarization slice, rather than complete the analysis pixel-by-pixel as is normally done for GPI data. This procedure avoids comparing pixels with greatly different signal to noise ratios as the PSF position and shape changes between images. For each datacube we summed the total flux in each slice and placed it into a single pixel in that slice. To calculate the Stokes vector associated with each waveplate sequence, the four polarization datacubes (8 single-value pixels) were combined using the Combine Polarization Sequence primitive in the DRP with the default settings. For observing configurations where multiple waveplate sequences were taken in a row we calculated a Stokes vector for each waveplate sequence. 

\begin{figure}
\begin{center}
\includegraphics[width=0.7\columnwidth]{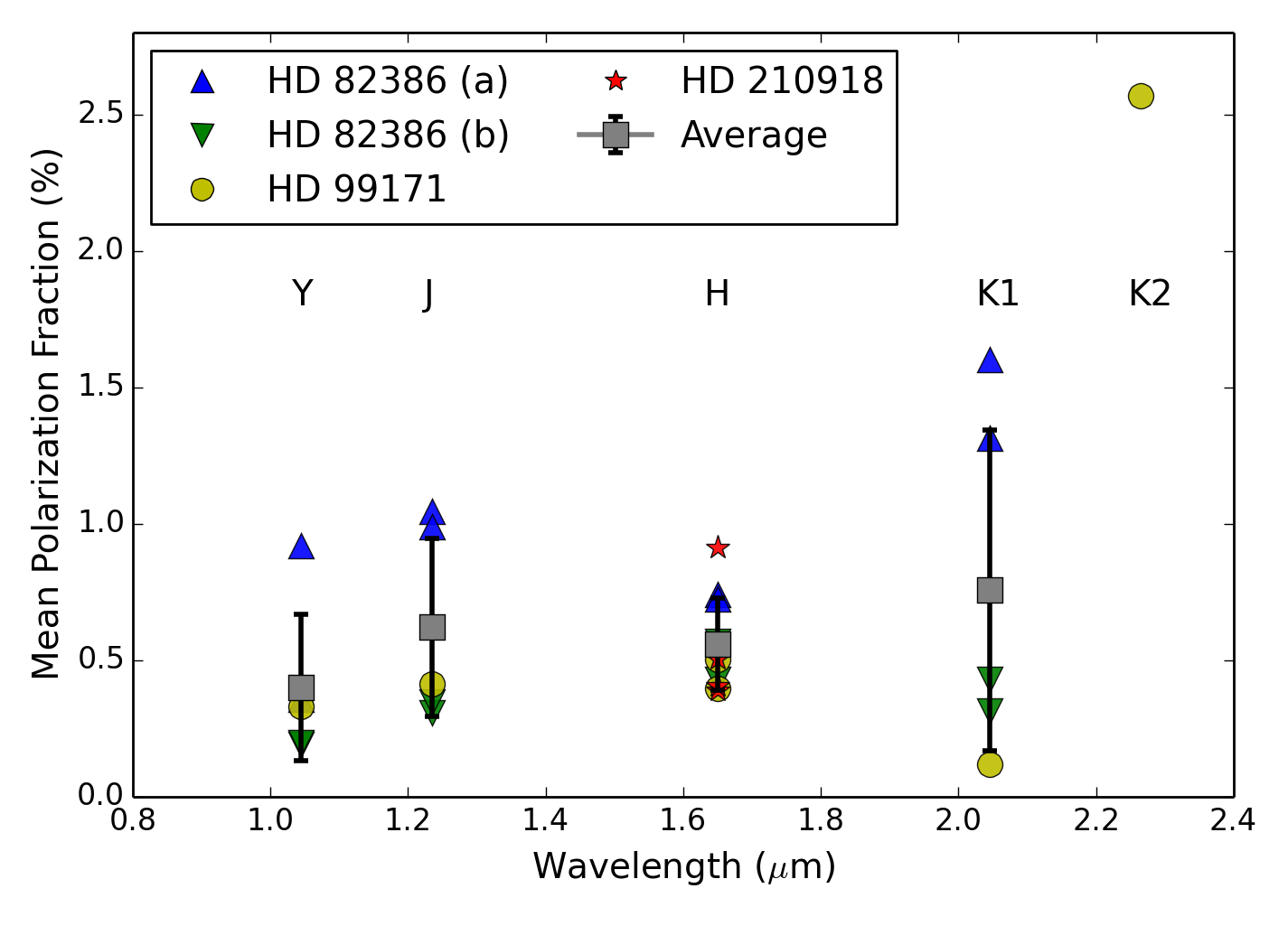}
\end{center}
\caption[Instrumental Polarization of GPI Estimated with Unpolarized Stars]{Polarization fraction measurements of unpolarized stars expressed as a percent. The different targets are denoted with different shapes and different colors. HD 82386 (a) and (b) data sets were taken 4 days apart. The $K2$ measurement may have been corrupted due to a misaligned polarization calibration solution. The mean polarization fractions for $Y$, $J$, $H$, and $K1$ across all observations are shown as grey squares, with error bars representing the sample standard deviations. The instrumental polarization appears to be color-independent within errors, with an average of $\sim 0.6\%$.
\label{fig:unpolstd}}
\end{figure}

The linear polarized fraction from each Stokes vector can be seen in Figure~\ref{fig:unpolstd} as a function of wavelength. Among all the observations, there is only one waveplate sequence taken in the $K2$ band. The polarization datacube images of this $K2$ data set show a Mori\'e pattern, which is a known artifact from the data pipeline procedure caused by a misalignment between the polarization spot locations in the polcal and the data. Unfortunately, simply offsetting the calibration solution or using the standard flexure correction procedure could not mitigate this problem. We therefore consider that this measurement may be spoiled. 

\begin{table}
\caption[Unpolarized Standard Stars Observed in Direct Mode as Polarimetric Calibrators.]{Unpolarized standard stars observed in direct mode as polarimetric calibrators. Filters are listed in the chronological order of the observations. Each sequence consists of 4 images, where the waveplate was rotated by $22.5^{\circ}$ in between each exposure. The exposure times are for each individual exposure, not the sequences as a whole.} 
\label{tab:unpol}
\begin{center}       
\begin{tabular}{|c|c|c|c|c|} 
\hline
\rule[-1ex]{0pt}{3.5ex}  Star Name & UT Date & Filter Sequence & \# of HWP Sequences per Filter &  Exp. Time (s) \\
\hline
\rule[-1ex]{0pt}{3.5ex}  HD 82386 & 2014-03-21 & $H$, $J$, $Y$, $K1$ & 2 & 12 \\
\hline
\rule[-1ex]{0pt}{3.5ex}  HD 82386 & 2014-03-25 & $K1$, $H$, $J$, $Y$ & 2 & 12 \\
\hline
\rule[-1ex]{0pt}{3.5ex}  HD 99171 & 2014-05-12 & $H$, $J$, $Y$, $K1$, $K2$, $H$ & 1 & 30 \\
\hline
\rule[-1ex]{0pt}{3.5ex}  HD 210918 & 2014-09-10 & $H$ & 4 & 15 \\
\hline
\end{tabular}
\end{center}
\end{table} 

All of the stars observed in our sample are unpolarized standard stars and we consider the detected polarization signals due to instrumental polarization. The percentage (\%) mean polarization fractions for $Y$, $J$, $H$, and $K1$ are $0.40 \pm 0.27$, $0.62 \pm 0.37$, $0.56 \pm 0.17$, and $0.76 \pm 0.59$ respectively. The uncertainties represent the sample standard deviations. We find our $H$-band observation to be consistent with the measurement made by Wiktorowicz et al.\cite{Wiktorowicz2014SPIE} using $\beta$ Pic. However, all of our observations are shorter sequences and have less field rotation than the observations of $\beta$ Pic and as a result the errors are larger. Our results indicate that the instrumental polarization appears to be color-independent to within our errors. This is consistent with the analysis presented by Wiktorowicz et al.\cite{Wiktorowicz2014SPIE} who use their $H$-band measurement to calibrate GPI $J$, $H$ and $K1$ observations of the polarized stars HD 77581 and HD 78344. Using this calibration they successfully fit a Serkowski law to a combination of their GPI measurements and previous visible light measurements. 

We note that noise in the measurements of $Q$ and $U$ can introduce a bias when calculating the mean polarization fraction due to the squared $Q$ and $U$ terms in the calculation of $P$. Even when both $Q$ and $U$ measurements have zero means, the means of $Q^2$ and $U^2$ will not be zero as long as the sample standard deviations of $Q$ and $U$ are not zero. This effect will introduce a positive bias in the mean of $P$, with the magnitude of the bias dependent on the sample standard deviations of $Q$ and $U$. We take a numerical approach to estimate this bias. For each band, we draw two large random samples: one from a normalized Gaussian with the mean of zero and $\sigma$ of the sample standard deviation of $Q/I$, and the other with the same Gaussian except for having the $\sigma$ being the sample standard deviation of $U/I$. We then use those values to calculate $P$. By taking the average of $P$, we estimate the bias in the polarization fraction in percentage (\%) for $Y$, $J$, $H$, and $K1$ to be 0.29, 0.30, 0.37, and 0.56, respectively. All of these biases are lower than the $P$ measurements of the unpolarized standard stars, showing the presence of the instrumental polarization over this bias. Due to the varying PSF during the open loop observations we consider the variation in the sample's $Q/I$ and $U/I$ to represent an upper limit in the uncertainty in $Q$ and $U$ and so these bias estimates are considered to be strong upper limits. 


\begin{figure}
\begin{center}
\includegraphics[width=0.7\columnwidth]{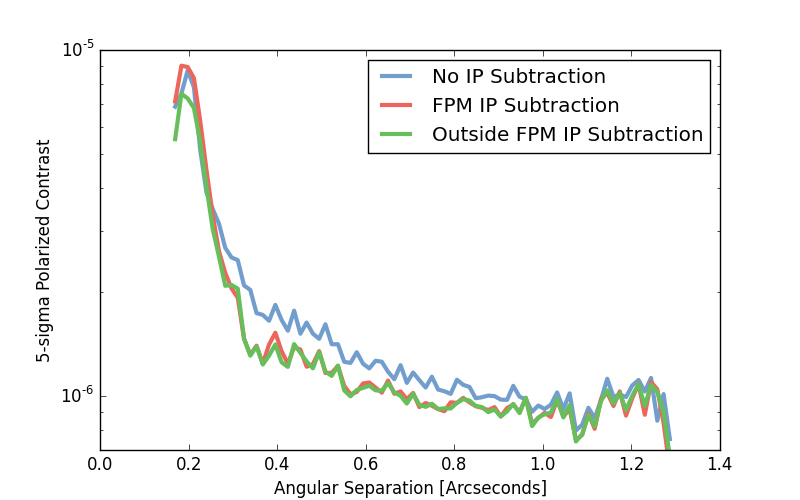}
\end{center}
\caption[Polarized Intensity Contrast Using Different Instrumental Polarization Subtraction Techniques]{Linear polarized intensity 5-sigma contrast for the observations of HD 118666 as a function of angular separation using different instrumental polarization subtraction techniques. For this dataset, when the instrumental polarization (IP) is estimated using flux measured behind the FPM, gains in contrast are seen between $\sim 0.3$ and 1.0 arcseconds. Further contrast gains can be achieved at smaller separations when the instrumental polarization is measured using flux from just outside the FPM. Note that when there is significant astrophysical flux at small angular separations, measuring the instrumental polarization outside of the FPM runs the risk of subtracting real polarized signal from the datacubes.  
\label{fig:ip_contrast}}
\end{figure}

\subsection{Subtracting Apparent Stellar Polarization from Polarization Datacubes}
\label{sec:sub_ip}
In a standard coronagraphic observing sequence the instrumental polarization can be subtracted in each polarization datacube by measuring the apparent stellar polarization. The apparent stellar polarization is measured as the mean fractional polarization at the location of the focal plane mask (see Figure~\ref{fig:ip_demo}), and contains contributions from the instrumental polarization, interstellar polarization and possibly polarized scattered light on angular scales less than GPI's diffraction limit. We expect that for most of GPI's targets the instrumental polarization will be the dominant term. The fractional polarization is defined as the difference of flux between the two orthogonal polarization slices divided by the total flux in both slices. Any light in this area will be light that has diffracted around the FPM and should be almost entirely due to the light of the star (a similar effect as an Arago or Poisson spot).  The polarized flux caused by the apparent stellar polarization at a given spatial location can then be estimated by scaling the measured fractional polarization behind the coronagraph by the total flux at that location. This signal can then be subtracted out from each datacube individually. This method has been implemented in the most recent release of the GPI DRP as a primitive called ``Subtract Mean Stellar Polarization" and has been used in a number of recent GPI publications \cite{MillarBlanchaer2015, Kalas2015, Draper2016}. 

To demonstrate the effects of this procedure we applied it to the GPI commissioning observations of the unpolarized standard star HD 118666 discussed in Section~\ref{sec:extraction}. Figure~\ref{fig:ip_contrast} displays the improvement in linear polarized intensity contrast in the final combined Stokes cube when subtracting the instrumental polarization. Noticeable improvements are realized between $\sim$0.3 and 1 arcseconds. Figure~\ref{fig:qrur_ip_sub} displays $Q_r$ and $U_r$ images (where the effects of apparent stellar polarization polarization are most apparent) of HD 118666, before and after subtraction. Though gains are made in the inner regions, it appears that we are still limited by systematics rather than random noise inside of ~0.25$''$. 

In some cases (e.g. for very faint stars or very short exposure times) the amount of flux behind the FPM may be extremely low, resulting in a poor S/N estimate for the apparent stellar polarization. In this case, it may instead be estimated using the light just outside of the coronagraph, where the stellar flux is the highest (Figure~\ref{fig:qrur_ip_sub}). The increased flux in this area results in a higher S/N measurement. However, caution must be exercised when using this region to measure instrumental polarization; if there is a highly polarized source near the edge of the FPM, the assumption that the measured fractional polarization is due solely to instrumental/stellar polarization may break down. The ``Subtract Mean Stellar Polarization" gives the user the ability to choose from which area they wish to measure the instrumental polarization, with the default being behind the FPM. 

The general strategy of subtracting the instrumental polarization from each polarization datacube individually has several advantages. First, if the host star exhibits some level of polarization, we can use this method to measure and subtract the stellar polarization that may dilute any measurement of polarized circumstellar material. However, stellar polarization is typically due to polarization from interstellar dust grains, and the magnitude of the signal increases with distance from the earth\cite{Fosalba2002}. The majority of targets that are appropriate for observation with GPI are nearby and should have negligible interstellar polarization. A second advantage is that this method is robust against a changing instrumental polarization, that may vary with time or telescope elevation. In fact, by recording the fractional polarization measured in each frame we are able to monitor and track any changes in the instrumental polarization over time. The analysis of this data is ongoing and will be published at a later time. 


\section{GPI Exoplanet Survey Contrasts}
\label{sec:GPIES}

The GPIES campaign is a multi-year Gemini South program with the goal of discovering and characterizing directly imaged exoplanets around young nearby stars using GPI's spectroscopy mode. The campaign also includes a debris disk component with the goal of imaging and characterizing debris disks using GPI's polarimetry mode. Debris disk observations are split into two categories: a shorter snapshot sequence, whose purpose is detecting disks previously unseen in scattered light; and a deeper observation sequence, to obtain higher S/N data for detailed disk characterization. The exoplanet search will target a total of 600 stars in spectroscopy mode and those with a known infrared excess are observed as a polarimetric snapshot immediately following the spectroscopic observations. In total there are roughly 60 targets that will receive a snapshot observation. 

\begin{figure}
\begin{center}
\includegraphics[width=0.7\columnwidth]{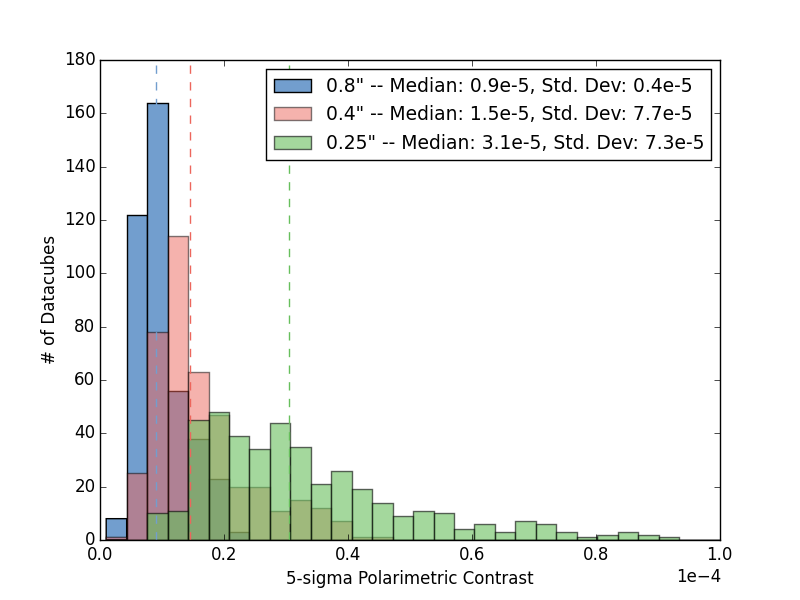}
\end{center}
\caption[Polarization Datacube Contrasts from GPIES]{A histogram of polarized intensity contrasts from single polarization datacubes obtained throughout the GPIES survey, measured at separations of 0.25$''$(green), 0.4$''$(red) and 0.8$''$(blue). The dotted lines represent the median of each distribution. The sample has been culled of any targets that have detected debris disks.
\label{fig:GPIES_podc_contrasts}}
\end{figure}

A typical polarimetry snapshot is between $8$ and $16$ 1-minute observations, with the HWP rotating between each. At the beginning of the campaign it was standard to obtain 8 1-minute observations for a snapshot. However, because polarimetric snapshots are obtained immediately after a spectroscopic sequence, persistence can significantly impact the depth of the observation sequence (Section~\ref{sec:persistence}). As a result, the length of a snapshot was first increased to 12 frames and as of Jan 2016 the standard snapshot time was further increased to 16 frames. Disks detected in a snapshot, or those previously resolved in scattered light are observed with a deep polarimetric sequence, typically 40 1-minute observations. In practice, telescope tracking errors and/or the opening of the adaptive optics control loops can lead to one or more frames being unusable. Thus, the exact number of frames used when forming a Stokes datacube can vary, but is nonetheless on the order of 16 and 40, for snapshot and deep observations, respectively. 

As of 2016 May 19, GPIES has observed 24 polarimetric snapshots and 18 deep sequences,  amounting to 861 polarization datacubes and 42 Stokes datacubes. Figure~\ref{fig:GPIES_podc_contrasts} displays histograms of the polarized intensity contrasts at angular separations of 0.25$''$, 0.4$''$ and 0.8$''$ of all the polarization datacubes without detected disks. The distributions at 0.4$''$ and 0.8$''$ show strong peaks near a contrast of 1e-5, with sharp drops to smaller values and small tails that trail off to higher values. This likely indicates that we are reaching the photo/read out noise boundary at these separations for most of our observations, as suggested in Section~\ref{sec:extraction}. On the other hand, the distribution for 0.25$''$ is much broader. At these separations we believe we are limited by instrumental polarization and our ability to remove it, though future tests will confirm this. Note that in this plot the datacubes have neither been cleaned with the double differencing process nor had instrumental polarization subtracted, because the contrast is typically measured before the cubes are combined in any way. Thus, the values shown here can be considered upper limits. 

\begin{figure}
\begin{center}
\includegraphics[width=0.9\columnwidth]{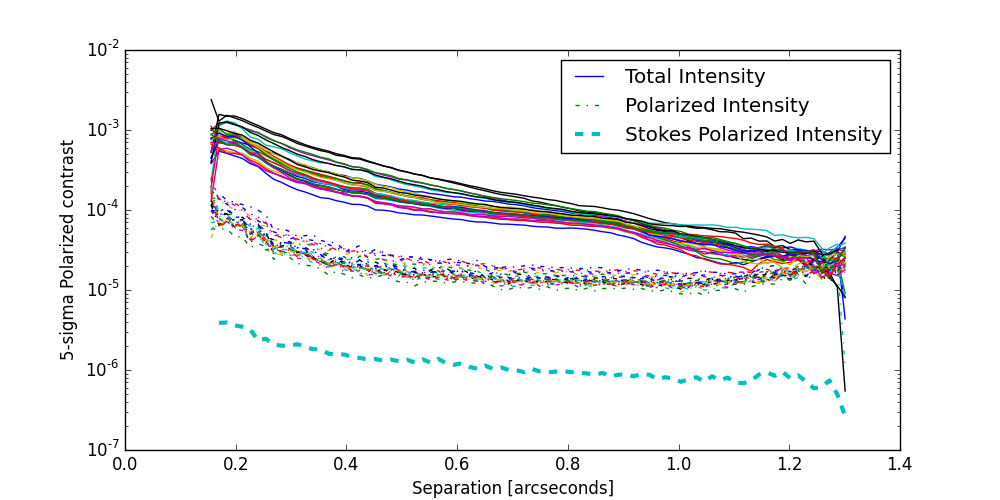}
\end{center}
\caption[Contrasts from GPIES Observations of HD 10472]{Contrasts from the GPIES observations of HD 10472. Displayed at the top are the total intensity and linear polarized intensity contrasts from each polarization datacube for the entire observation set (26 1.5-minute exposures). The bottom line displays the linear polarized intensity contrast of the combined Stokes datacube. 
\label{fig:deep_pol}}
\end{figure}

\begin{figure}
\begin{center}
\includegraphics[width=0.7\columnwidth]{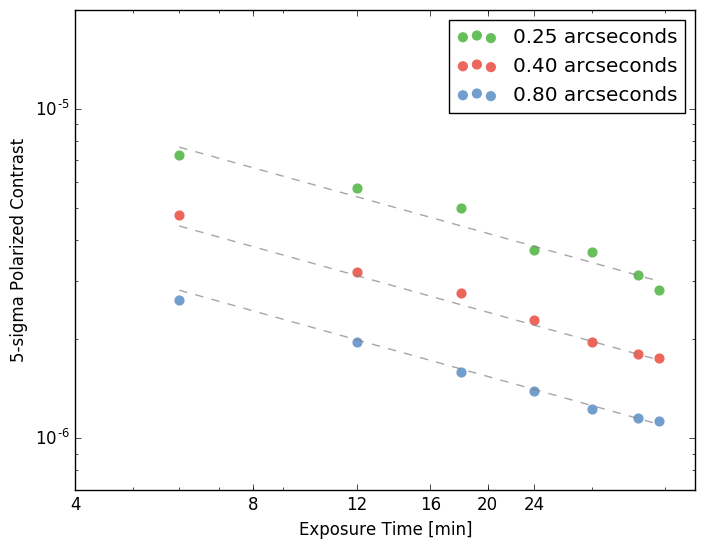}
\end{center}
\caption[Contrast as a Function of Exposure Time for HD 10472]{Polarized contrast at 0.25$''$, 0.40$''$ and 0.80$''$ as a function of total exposure time for the HD 10472 dataset. The polarized contrast at all three separations appears to decrease as a function of $\sqrt{\text{Exposure Time}}$ (grey dashed lines). 
\label{fig:HD10472_contrast_time}}
\end{figure}

\begin{figure}
\begin{center}
\includegraphics[width=\columnwidth]{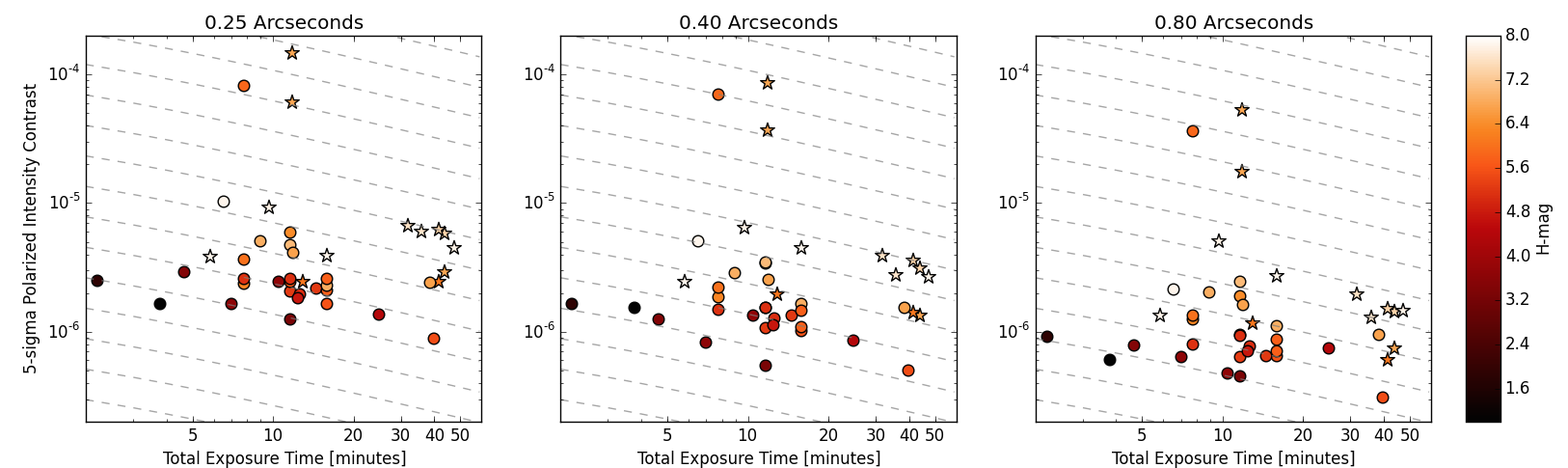}
\end{center}
\caption[Polarized Intensity Contrasts at 0.25$''$, 0.4$''$ and 0.8$''$ for all GPIES Disk Observations]{Polarized intensity contrasts at 0.25$''$, 0.4$''$ and 0.8$''$ from all the observations throughout the GPIES survey as a function of number of exposures and H-magnitude. Each exposure consists of a 60-s observation, so the x-axis can be considered a proxy for exposure time. Observations with detected disks are marked with a star symbol, and non-detections are marked with circles. The grey dashed lines indicate contrasts decreasing as $\sqrt{\text{Exposure Time}}$.
\label{fig:GPIES_stokesdc_contrasts}}
\end{figure}

In Figure~\ref{fig:deep_pol}, we display contrast curves from observations of HD 10472, as an example of a typical GPIES deep polarization observing sequence. HD 10472 was observed on 2015 December 12, as part of the GPIES campaign and the observation set consist of 26 1.5-minute exposures. The figure simultaneously displays the total intensity contrast and polarized intensity contrast of the polarization datacubes, as well as the linear polarized intensity contrast of the final Stokes cube. The median total intensity contrasts of the polarization datacubes at 0.25$''$, 0.4$''$ and 0.8$''$ are \num{4.2e-4}, \num{1.7e-4} and \num{0.8e-4}, respectively. The median linear polarized intensity contrasts of the polarization datacubes at 0.25$''$, 0.4$''$ and 0.8$''$ are \num{0.44e-4}, \num{0.19e-4} and \num{0.12e-4}, respectively, providing an improvement of roughly a factor of 10 at 0.25$''$ and 0.4$''$, and a factor of 6.5 at 0.8$''$. The linear polarized intensity contrasts of the final Stokes cube at 0.25$''$, 0.4$''$ and 0.8$''$ are \num{0.024e-4}, \num{0.015e-4} and \num{0.010e-4}, respectively. Thus, we gain factors of roughly 175, 110 and 80 between the total intensity and final linear polarized intensity at 0.25$''$, 0.4$''$ and 0.8$''$, respectively. In this observation set, we find that polarized contrast at all three separations decreases as a function of $\sqrt{\text{Exposure Time}}$ (Figure~\ref{fig:HD10472_contrast_time}). 

Final contrasts for all the Stokes cubes produced so far in GPIES (both snapshot and deep observations) can be seen in Figure~\ref{fig:GPIES_stokesdc_contrasts} as a function of exposure time and H-magnitude. The plot includes observations both with non-detections and with detected disks. Because the disk flux likely increases the measured contrast, contrasts associated with detected disks can be considered upper limits on the true sensitivity of the observations. In general, deeper contrasts are achieved for brighter targets, and targets with the same H-magnitude appear to gain in contrast roughly as $\sqrt{\text{Exposure Time}}$  (with a few exceptions).

\section{Conclusions}
\label{sec:conclusions}

The Gemini Planet Imager has now been on-sky for over 2.5 years and is producing exciting results through the GPIES campaign, the debris disk Large and Long Program and through Gemini queue observations. Over this time we have developed new data analysis techniques that we have implemented as part of the publicly available GPI DRP. 

Three techniques in particular have allowed us to reduce systematics and improve upon the contrast in GPI's Polarimetry mode. First, by using a weighted PSF extraction in the assembly of polarization datacubes from raw data we were able to lower the photon noise/read noise floors of the polarization datacubes, which improves the contrast between 0.3$''$ and the edge of the field. Second, by applying a polarized flat-field we can reduce systematics near the edge of the field. Third, by measuring and subtracting the apparent stellar polarization we can subtract the instrumental polarization, with contrast benefits from the inner working angle to about 1$''$. All of these improvements have been included in the GPI DRP and are used in standard reduction recipes of GPIES data. In addition we perform a multi-wavelength analysis of GPI's instrumental polarization, that indicates that the instrumental polarization appears to be roughly wavelength independent. 

The combined datasets of the GPIES campaign demonstrate that the polarized contrast can be improved with increased exposure time and the achievable contrast depends on the brightness of the source. An examination of histograms of the contrast of polarization datacubes and the $\sqrt{\text{Exposure Time}}$ dependence of the contrast of a deep polarization sequence indicate that sensitivity at 0.4$''$ and 0.8$''$ is likely dominated by photon/read noise (with small contributions from instrumental polarization). We believe to be dominated by residual instrumental polarization or other polarization systematics at separations smaller than 0.25$''$. Nonetheless, we find that the polarized contrast decreases as a funtion of $\sqrt{\text{Exposure Time}}$.  The best contrasts achieved so far as part of the campaign at 0.25$''$, 0.4$''$ and 0.8$''$ are \num{9e-07}, \num{5e-07} and \num{3e-07} respectively, obtained with 40 60s frames on a star with a 5.5 H magnitude. These measurements and others presented throughout this work can be used as baseline estimates of GPI's polarimetry mode's performance when planning future observations. 

A full characterization of the different noise contributions and systematics when using the weighted PSF subtraction technique has been left for future work. Ongoing work to further improve our sensitivity includes the development of methods to subtract persistence in raw data and to subtract instrumental polarization at small inner working angles.


\acknowledgments 
JRG was supported in part by NASA/NNX15AD95G. 

\bibliography{spie} 
\bibliographystyle{spiebib} 

\end{document}